\documentclass[]{emulateapj}
\usepackage{graphicx}

\def\etal{{et al.}}
\def\ie{{i.e.}}
\long\def\Ignore#1{\relax}
\usepackage{color}
\definecolor{red}{rgb}{0.7,0.1,0.1}

\shorttitle{Bar slowdown}
\shortauthors{Sellwood \& Debattista}

\begin{document}

\title{Re-interpretation of ``Bar slowdown and the distribution of \\
  dark matter in barred galaxies'' by Athanassoula}

\author{J. A. Sellwood}
\affil{Department of Physics \& Astronomy, Rutgers the State
  University of New Jersey, 136 Frelinghuysen Road, Piscataway, NJ
  08854, USA}

\author{Victor P. Debattista}
\affil{Jeremiah Horrocks Institute, University of Central Lancashire,
  Preston, PR1 2HE, UK}

\begin{abstract}
Athanassoula (2014) has claimed that measurements of the ratio of
corotation radius to bar length in galaxies do not place a constraint
on the disk to halo mass ratio.  Her conclusion was based on a series
of simulations published by \citet{AMR13}.  Here we show that these
results are, in fact, consistent with previous work on the slow down
of bars due to dynamical friction because gas inflow rearranges the
disk mass before the bar settles.  It therefore remains true that a
bar rotating sufficiently fast that corotation is not far beyond the
bar end requires a near maximum disk.
\end{abstract}


\section{Context}
The issue of whether the baryonic material in galaxy disks does, or
does not, contribute most of the central attraction is important for
many reasons, such as understanding the dynamical structure of
galaxies and comparison with predictions from galaxy formation models.
While the total central attraction is determined by the rotation
curve, the relative contributions of baryonic and dark matter are not
easily separated.  Various indirect arguments have been advanced to
bear on this issue, one of which is that strong bars in sub-maximal
disks should be slowed by dynamical friction \citep{DS00}.  Since
$R_c$, the radius of corotation for a bar, increases as the bar slows,
we proposed an observationally accessible measure of whether a bar has
been slowed by friction as the value of the dimensionless ratio ${\cal
  R} = R_c/a_B$, where $a_B$ is the semi-major axis of the bar.  We
argued that a strong bar can remain fast, ${\cal R} \la 1.4$, only if
the barred disk is close to maximal.

However, this argument was called into question in a recent paper by
\citet{Atha14} who reported $\cal R$ values from a suite of 15
simulations with initial disk gas fractions ranging from 0 to 100\% in
halos that were axisymmetric, as well as similar models with mildly or
strongly triaxial halos.  She concluded ``The models, by construction,
have roughly the same azimuthally averaged circular velocity curve and
halo density and they are all submaximal, \ie\ according to previous
works they are expected to have all roughly the same ${\cal R}$ value,
well outside the fast bar range ($1.2 \pm 0.2$). Contrary to these
expectations, however, these simulations end up having widely
different ${\cal R}$ values, either within the fast bar range, or well
outside it.  This shows that the ${\cal R}$ value can not constrain
the halo density, nor determine whether galactic discs are maximal or
submaximal.''

Here we show that these statements reflect a simple misunderstanding
of the criterion, and that her results appear to be in good agreement
with it.

\begin{figure*}
\begin{center}
\includegraphics[width=.27\hsize, angle=270]{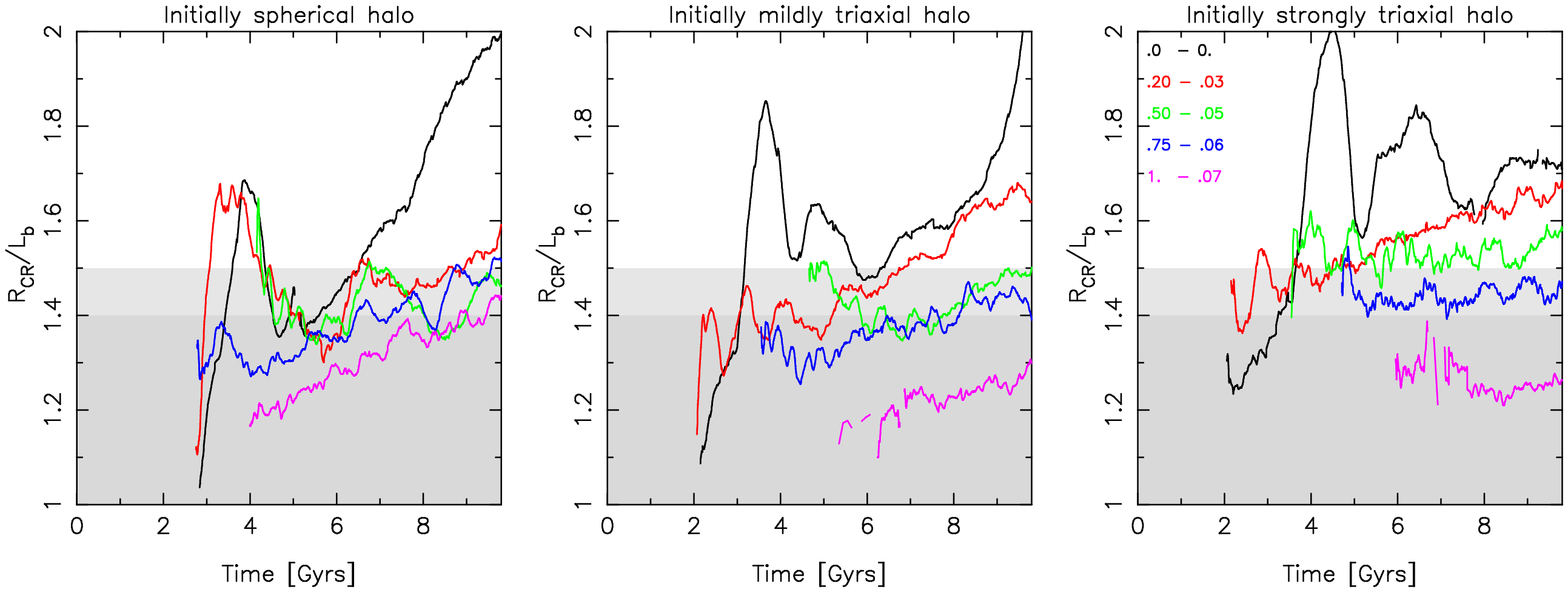}

\medskip
\includegraphics[width=.25\hsize, angle=270]{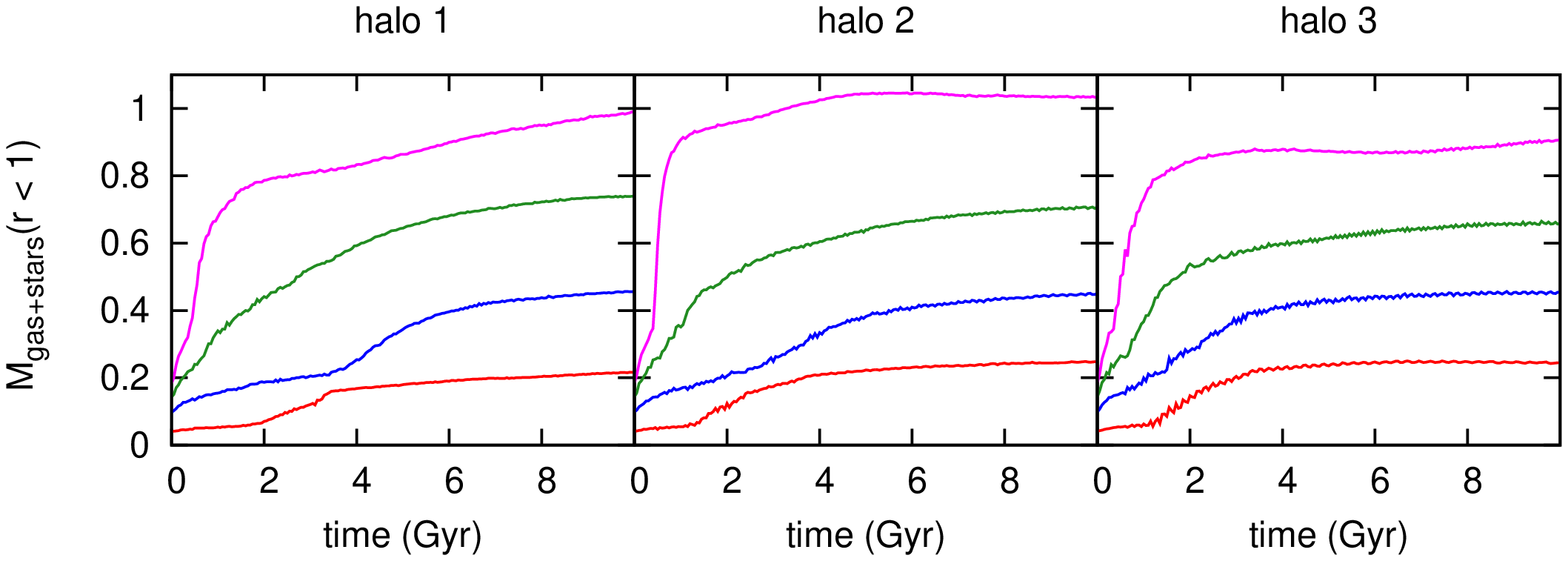}
\end{center}
\caption{Above, Figure 1 reproduced from \citet{Atha14}, showing the
  time evolution of her measurements of ${\cal R}$ in all 15 of her
  simulations.  Below, Figure 16 from \citet{AMR13} showing the time
  evolution of the gas+new star particles inside $R=1\;$kpc in the
  same simulations.  The red and magenta lines relate to the same
  models in both rows of figures, but the blue and green lines relate
  to models that have been interchanged between the two rows.}
\end{figure*}

\bigskip\bigskip\bigskip
\section{The evidence}
The claims by \citet{Atha14} were based on calculations that were
described more fully in \citet{AMR13}.  They all began with
identically the same mass distribution, except that the disk was
composed of different fractions of gas and star particles, and the
spherical halo in one set was distorted into mildly and more strongly
triaxial shape in the other two sets.  At the outset, the disk was
clearly submaximal, since the peak circular speed from the disk was
fractionally less than the circular speed from the halo at the same
radius.

The simulations lacking any gas, formed large ($a_B \sim 10\;$kpc),
strong bars that were fiercely braked by dynamical friction against
the halo, so that ${\cal R} \gg 1.4$ by the end of the simulation.  It
is clear from Fig.~1 of \citet{Atha14}, reproduced here in the upper
panel of our Fig.~1, that the final ${\cal R}$ values decreased
systematically with increasing gas fraction.

However, as shown in the lower panel of our Fig.~1, which is
reproduced from Fig.~16 of \citet{AMR13}, the mass distribution in the
disks with gas was rearranged within the first 2~Gyr of evolution,
well before the bars had formed and settled.  The mass that
accumulated in the inner disk increased with the gas mass fraction,
and in the cases that began with 100\% gas (magenta lines) the mass
within 1~kpc increased some 4-fold over that at the start.  The other
simulations start with smaller gas fractions, and larger stellar
fractions (the initial stellar fraction is omitted in their Figure),
and therefore the relative increase in the total inner disk mass is
proportionately less.

This initial rearrangement of the mass distribution is sufficient to
change the dynamical properties of the models.  For example, if the
100\% gas disk were to have contracted homologously to an exponential
disk of half the radial scale but the same total mass, then the
central surface density would increase 4-fold, have a peak in the
circular speed that is $\surd2$ times higher than before lying at a
radius that is half that of the original peak.  From Fig.~1 of AMR13,
that would mean the disk contribution reaches $\sim200\;$km/s at
$R=3.3\;$kpc, which would clearly be maximal.

The contraction is unlikely to be homologous, but this crude
approximation is perhaps consistent with the initially gas-rich models
forming shorter bars.  The size of the bar that forms in a simulation
is determined by a variety of factors, including the steepness of the
inner rise in the rotation curve and the distribution of mass in the
disk.  The bar in the model with 100\% gas has $a_B \sim 5\;$kpc in
the spherical halo \citep[][her Fig.~2]{Atha14} about half that in the
stars only case, and the decreasing bar sizes can also be seen in the
snaphots of Figs.~4 \& 5 and the measurements in Fig.~9 of
\citet{AMR13}.

It is therefore no surprise that the shorter bars in the increasingly
dominant disks experience weaker friction, such that ${\cal R}$
manifests the trend shown in the upper panel of Fig.~1.  It seems very
likely that the bars that experience little frictional drag are
effectively in maximum disks.

Note also that the initial rise in the central gas density was greater
in the models with triaxial halos.  The principal consequence of
triaxiality is to promote the inflow of gas, which is borne out by the
mass increase by $t=2\;$Gyr in all cases except for the 100\% gas
case, which is less in the strongly triaxial case than in the mildly.
This odd result could simply be stochasticity \citep{SD06} that could
be checked in multiple runs with different random seeds.

\section{Conclusions}
Once the mass rearrangement, due to the initial shrinking size of the
gas component, is taken into account, the simulations reported by
\citet{AMR13} and \citet{Atha14} appear to be consistent with all
previous work \citep[see][for a review]{Sell14}.  The different $\cal
R$ values at the ends of her 15 simulations can be understood as
reflecting the differences in the effective degree of disk domination
in the inner parts of the models.  Thus it would appear that her
results in agreement with the conclusion that only maximal disk models
can have ${\cal R} \sim 1.2 \pm 0.2$ after some period of evolution.

The statements to the contrary in \citet{Atha14} seem to reflect a
simple misunderstanding by that author of what exactly is the
criterion.  In fact, it relates to the mass distribution in the model
at the time of the measurement, which is the only observationally
accessible measurement, of course.  Her mis-statements derive from an
incorrect expectation that models that begin from the same mass
distribution, but with differing fractions of gas, should all
experience the same dynamical friction; in fact, the mass
rearrangement in the early part of the evolution makes the gas-rich
disks more nearly maximal, leading to correspondingly weaker friction
on the bar that subsequently forms.

\end{document}